\documentclass{article}
\usepackage{arxiv}

\usepackage[T1]{fontenc}    
\usepackage{hyperref}       
\usepackage{url}            
\usepackage{microtype}      
\usepackage{graphicx}

\title{Temperature-dependent optical properties of $\epsilon$-Ga\textsubscript{2}O\textsubscript{3} thin films}

\author{\href{https://orcid.org/0000-0002-5706-8909}{\includegraphics[scale=0.06]{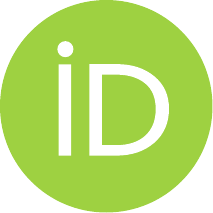}\hspace{1mm}Takayuki Makino} \\
	Far-IR R\&D Center\\
	University of Fukui\\
	Fukui 910-8507, Japan \\
	\texttt{tmakino@u-fukui.ac.jp} \\
\And
Subaru Yusa\\
Department of Chemistry,\\
Tohoku University\\
Sendai 980-8578, Japan\\
\And
\href{https://orcid.org/0000-0003-2747-9675}{\includegraphics[scale=0.06]{./img/orcid.pdf}\hspace{1mm}Daichi Oka}\\ 
Department of Chemistry,\\
Tohoku University\\
Sendai 980-8578, Japan\\
\And
\href{https://orcid.org/0000-0002-8957-3520}{\includegraphics[scale=0.06]{./img/orcid.pdf}\hspace{1mm}Tomoteru Fukumura}\\
Department of Chemistry and WPI-AIMR\\
Tohoku University\\
Sendai 980-8577, Japan\\
}

\renewcommand{\shorttitle}{Optical properties of $\epsilon$-Ga\textsubscript{2}O\textsubscript{3}}

\hypersetup{
pdftitle={optical properties of $\epsilon$-Ga\textsubscript{2}O\textsubscript{3}},
pdfsubject={cond-mat},
pdfauthor={T.~Makino},
pdfkeywords={Jpn. J. Appl. Phys.},
}

\begin{document}
\maketitle

\begin{abstract}
We determined the complex dielectric functions of
$\epsilon$-Ga\textsubscript{2}O\textsubscript{3} using optical transmittance and
reflectance spectroscopies at temperatures from 10 K to room
temperature. The measured dielectric-function spectra reveal distinct
structures at bandgap energy. We fitted a model dielectric function
based on the electronic energy-band structure to these experimental
data. We analyzed the temperature dependence of the bandgap with a model
based on phonon dispersion effects. One could explain it in terms of
phonon-related parameters such as the optical phonon temperature. We
compare phonon-related properties of
$\epsilon$-Ga\textsubscript{2}O\textsubscript{3} with those of a large variety of
element and binary semiconductors.
\end{abstract}

\hypertarget{introduction}{%
\section{1. Introduction}\label{introduction}}

Gallium oxide (Ga\textsubscript{2}O\textsubscript{3}) is an important
wide-gap material from both fundamental and applied perspectives due to
its potential applications such as power devices\textsuperscript{1),2)}
and solar-blind UV photodetectors\textsuperscript{3),4)}. There are
several polymorphs in this oxide.\textsuperscript{2)} The most stable
phase among them is
\(\beta\)-Ga\textsubscript{2}O\textsubscript{3}\textsuperscript{1),4)}.
This polymorph has an indirect bandgap of 4.4--4.9
eV\textsuperscript{5)} and monoclinic crystalline
symmetry.\textsuperscript{1)} Despite intensive studies conducted on
this polymorph\textsuperscript{5)} \textsuperscript{6)}, its complicated
lattice structure became a bottleneck of further progress, such as the
growth of heterojunctions with other materials\textsuperscript{7),8)}.
On the other hand, corundum $\alpha$-Ga\textsubscript{2}O\textsubscript{3} has
a relatively higher symmetry of the crystal\textsuperscript{9),10)}. And
thus, an epitaxial thin film growth as a semiconductor device and its
heteroepitaxial structures have been reported so
far\textsuperscript{11),12)}. It is also possible to grow
$\epsilon$-Ga\textsubscript{2}O\textsubscript{3}\textsuperscript{13),14)}.

After pioneering work on the epitaxial thin film growth of
$\epsilon$-phase,\textsuperscript{15)} many research groups have reported various
studies on this polymorph\textsuperscript{16),17)}. The X-ray
diffraction studies initiated the long-standing debate on the
crystalline structure of this polymorph\textsuperscript{17),18)}. Cora
and coworkers\textsuperscript{18)} reported the
ordered-oxygen-vacancy-site-induced reconstruction into the orthorhombic
form\textsuperscript{19),20)}, sometimes referred to as the $\kappa$-phase.
Attempts on the growth onto various substrates followed this seminal
work\textsuperscript{21),22),23)}. More recently, a substrate-dependent
properties assessment further developed systematical insight into this
polymorph's potential possibilities\textsuperscript{24),25),26)}.

Furthermore, the recent reports on the high dielectric
constants\textsuperscript{27)} and the suitability of the
heterostructure growth\textsuperscript{24)} give rise to
polymorph-dependent characteristic properties that may exhibit solely in
this polymorph. However, further investigation of its basic material
properties is necessary to realize its full potential. The temperature
dependence of the optical properties is one of a semiconductor's
fundamentally and technologically essential
characteristics\textsuperscript{28),29)}. One may expect that this
investigation offers more profound insights into various aspects such as
electron-phonon interaction, thermal properties, and phononic
properties. So far, spectroscopic measurements have been done only at
room temperature\textsuperscript{23),27)}. In other words, no one has
conducted this in the cryogenic temperature range for this polymorph to
the best of our knowledge.

In this paper, we report the temperature dependence of the optical
properties of $\epsilon$-Ga\textsubscript{2}O\textsubscript{3} thin films. We
emphasize the interpretation of the temperature dependence of the
bandgap energies from the viewpoint of the electron-phonon interaction
particularly\textsuperscript{30),31)}. Finally, we compare the phononic
properties of Ga\textsubscript{2}O\textsubscript{3} with those of other
materials.

\hypertarget{experimental-methods}{%
\section{2. Experimental methods}\label{experimental-methods}}

$\epsilon$-Ga\textsubscript{2}O\textsubscript{3} epitaxial thin films were grown
at 570 \textsuperscript{o}C on sapphire (0001)
substrates\textsuperscript{27),32)}. Our adopted technique is mist
chemical vapor deposition (CVD). This technique is a facile solution
process at atmospheric pressure.\textsuperscript{12)} In its growth, one
used an aqueous solution of 0.025 M gallium acetylacetonate and 0.05 M
hydrochloric acid as a precursor. We nebulized the precursor solution
using ultrasonic transducers. We transferred it to the substrates in a
tubular furnace with nitrogen gas flow from the carrier and dilution
ports at 3.0 L/min and 0.5 L/min, respectively. The crystalline
structure of this phase and its structural properties can find
themselves in Figs. 1 and 2 of Ref. 27, indicating the high crystalline
quality of our films.

We measured the transmission (\(T\)) and reflection (\(R\)) spectra at a
beamline of the UVSOR facility in the Institute of Molecular Science,
Japan. One attached the samples on a cold finger of the refrigerator for
the temperature-dependent measurements. We do not plot the transmission
or reflection spectrum but the dielectric functions because of the
intimacy with various electronic-structure-related quantities. First, we
modeled our Ga\textsubscript{2}O\textsubscript{3}/sapphire system to
determine the real and imaginary components of dielectric
constants.\textsuperscript{33)} We used a homemade multilayer analysis
program to do this task. The complicated conversion is necessary between
optical and dielectric functions because these are related to both the
film and substrate parameters. In addition, the equations giving the
reflectivity and transmission explicitly in terms of the dielectric
constants of the film and the substrate are very complicated and have
multiple solutions. An iterative Newton-Raphson method determines the
dielectric constants by minimizing an error function composed of
calculated and experimental values. Denton and
coworkers\textsuperscript{33)} showed that using expressions for
\(\left( 1 \pm R \right)/T\) rather than the explicit formulae for the
reflectivity and the transmission simplifies the problem and eases the
computing burden. In this way, we found those components that best
explain the experimental data. We performed this analysis sequentially
for each wavelength to obtain the dielectric functions.

\hypertarget{results-and-discussions}{%
\section{3. Results and discussions}\label{results-and-discussions}}

\hypertarget{relationship-with-electronic-structures-and-model-dielectric-function-analysis}{%
\subsection{3.1 Relationship with electronic structures and model
dielectric function
analysis}\label{relationship-with-electronic-structures-and-model-dielectric-function-analysis}}

For purposes of discussion, we consider the band structures of
Ga\textsubscript{2}O\textsubscript{3}\textsuperscript{34),35)}. The$\kappa$
related band diagram can find itself in Fig. 2 of Ref. 35. In many
semiconductors, optical transitions related to the fundamental
absorption edge associate with three-dimensional direct
\emph{M}\textsubscript{0}-type critical points\textsuperscript{36),37)}.
Because there has been no report on strong anisotropy in the valence and
conduction bands as far as Ga\textsubscript{2}O\textsubscript{3} is
concerned\textsuperscript{34),35),38)}. Thus we assign the fundamental
absorption edge to \emph{M}\textsubscript{0} transitions from the
highest valence band to the lowest conduction band at the \(\Gamma\)
point. The corresponding optical transition is labeled
\emph{E}\textsubscript{0}. As later described, we did not account for
the spin-orbit split component. To account for the experimental curves
over the entire photon energy range, it was also necessary to include
the influence of the \emph{E}\textsubscript{1}
transitions.\textsuperscript{35)} Furthm\"uller and
coworkers\textsuperscript{34)} theoretically reported a
higher-lying-gap-related structure in the dielectric functions in the
$\epsilon$-phase at \emph{ca.} 5 eV. Because the calculation based on the density
functional theory underestimates the bandgap energies, we assign this to
the \emph{E}\textsubscript{1} gap at \emph{ca.} 6.8 eV by considering
this redshift effect (approximately 2 eV). The energetic position is
even higher than the measured photon energy range (6.2 eV). This
transition is labeled \emph{E}\textsubscript{1} in this work.

\begin{figure}
    \caption{Real (squares) and imaginary (circles) parts of the
    dielectric functions of Ga\textsubscript{2}O\textsubscript{3} thin-film
    taken at 190 K. The symbols are the experimental data. The lines are
    calculated for real (solid) and imaginary (dashed) parts using Eqs.
    (1)-(3), and $\epsilon_{1\infty}$ = 2.5.}
    \includegraphics[width=0.7\textwidth]{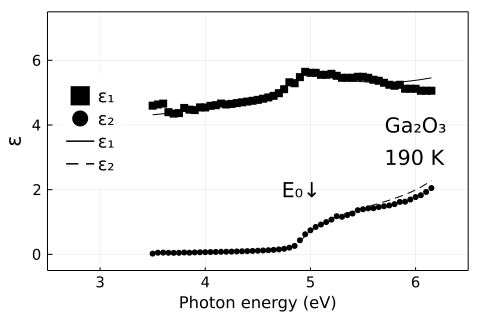}
    \end{figure}

We describe here the model dielectric functions. Transitions involving
these energy bands are responsible for all the features in the
dielectric functions. In a model that we have employed, one approximates
the dielectric function as a sum of several components, each of which is
an explicit function of energy. The dielectric function represents a
contribution from the neighborhood of a critical point in the joint
density of states. We use the following \emph{M}\textsubscript{0}-type
dielectric functions for the \emph{E}\textsubscript{0} transition. This
transition occurs at a photon energy of 4.89 eV at 190 K. Assuming the
bands are parabolic, we obtain the contribution of this gap to the
dielectric function\textsuperscript{36),37)}:

\begin{equation}
\epsilon\left( \hbar\omega \right) = \frac{A_{0}}{E_{0}^{1.5}}\left\lbrack \chi_{0}^{- 2}\left\{ 2 - \left( 1 + \chi_{0} \right)^{0.5} - \left( 1 - \chi_{0} \right)^{0.5} \right\} \right\rbrack,\ \#\left( 1 \right) \\
\end{equation}

With

\begin{equation}
\chi_{0} = \frac{\hbar \omega + i \Gamma}{E_{0}}
\end{equation}

Values of \(A_{0}\) and \(\Gamma_{0}\) in Eqs. (1) and (2) correspond to
the strength and broadening parameters in this optical transition.

We next describe the \emph{E}\textsubscript{1} feature. We labeled a
structure found in the optical spectra in the region higher than
\emph{E}\textsubscript{0} as \emph{E}\textsubscript{1} (\(\sim 6.8\)
eV). The \emph{E}\textsubscript{1} peak is difficult to analyze as it
does not correspond to single, well-defined critical points. Thus, we
have characterized the \emph{E}\textsubscript{1} structure (not
observed) as a damped harmonic oscillator\textsuperscript{36),37)}:

\begin{equation}
\epsilon\left( \hbar\omega \right) = \frac{A_{1}}{1 - \chi_{3}^{2} - i\Gamma_{1}\chi_{1}}\#\left( 3 \right) 
\end{equation}

with \(\chi_{1} = E/E_{1}\)Where \emph{A}\textsubscript{1} is a
\emph{dimensionless} strength parameter. The quantity \(\Gamma_{1}\)
stands for a \emph{dimensionless} broadening parameter.

We then describe the total dielectric function. The whole dielectric
function was found by summing the expressions given previously. We used
The parameters such as \emph{A}\textsubscript{0} and
\emph{A}\textsubscript{1} as adjustable constants to calculate the
dielectric functions. If we wish to obtain, for example, the imaginary
parts (\(\epsilon_{1}\)), we can take the imaginary part of Eqs. (1) and
(3). A constant, \(\epsilon_{1\infty}\) was added to the real part of
the dielectric constant (\(\epsilon_{1}\)) to account for the vacuum
plus contributions from higher-lying energy
gaps\textsuperscript{36),37),39)}. The model presented above can fit the
experimental dispersion of dielectric functions over the entire range of
the measurements.

Square and circle symbols in Fig. 1 show the real and imaginary parts of
the dielectric functions of Ga\textsubscript{2}O\textsubscript{3}. The
measurement temperature is 190~K. One obtained these data from the
analysis of the experimental data on \(T\) and \(R\). An arrow denotes
the value of the interband transition \emph{E}\textsubscript{0}. The
solid and dashed lines in Fig. 1 correspond to the results of the
least-squares fit. The model functions are Eqs. (1) - (3). There is
reasonable agreement between the experiment and the results of fit. We
evaluate \emph{E}\textsubscript{0}, \emph{A}\textsubscript{0}, and
\(\Gamma\)\textsubscript{0} to be 4.89 eV, 47~eV\textsuperscript{3/2},
and 45 meV, respectively. We can quickly determine an initial guess of
parameterization for the energy \emph{E}\textsubscript{0} because this
energy corresponds to the onset in the imaginary dielectric function
(\(\epsilon_{2}\)). Even after the regression analysis, this remained
almost unchanged. The obtained strength parameter value is reasonable
because it often ranges in tens of eV\textsuperscript{3/2} in many
semiconductors\textsuperscript{36),37)}. An empirical rule tells us that
the broadening parameter tends to be somewhat greater than the
corresponding thermal energy (\emph{ca.} 20 meV for 190 K), which is
also valid for this case.

Displayed in Figs. 2 and 3 are the individual contributions to the real
(\(\epsilon_{1}\)) and imaginary (\(\epsilon_{2}\)) parts of the complex
dielectric functions, respectively, of the two transitions. The related
equations are for the three-dimensional \emph{M}\textsubscript{0}
critical point and the \emph{E}\textsubscript{1} resonances. We evaluate
\(\epsilon_{1\infty}\) to be 2.5. Adachi and coworkers analyzed the
dielectric function to determine this value for many
semiconductors\textsuperscript{36),37)}. In many cases, this value is
often slightly larger than unity. The value obtained in this work is
significantly greater than that in other semiconductors. This
observation is consistent with the fact that the $\epsilon$-phase has a higher
static dielectric constant than the different phases of
Ga\textsubscript{2}O\textsubscript{3} or many semiconductors. Readers
might think the influence of strain disturbs the precise evaluation of
the bandgap energy. Indeed, there is strain between the substrate and
the thin film due to the lattice mismatch. We can safely rule out that
apprehension. Understanding the following facts can convince of its
validity. In general, only very high-precision spectroscopy can detect
very tiny energy shifts induced by externally applied stress. Due to the
relaxation, the lattice-mismatch-induced strain is considerably smaller
than the strain usually used in external stress experiments. Therefore,
we can neglect its influence on the bandgap energy. Likewise, we argue
the effect of inhomogeneity on the bandgap energy. We know well that the
inhomogeneity induces an exponential tail to the onset in the imaginary
dielectric function. We do not observe, however, such a tail state.
Therefore, we can safely rule out that possibility. The direct gap does
not exhibit a well-defined excitonic structure either. This absence is
probably consistent with the high-dielectric nature of this thin
film.\textsuperscript{27)} According to Go\~ni and
coworkers,\textsuperscript{40)} the binding energy of the exciton
includes the terms of the static dielectric constant and the effective
masses of the relevant electrons and holes. In other words, a higher
dielectric constant leads to smaller binding energy, while higher
effective mass gives rise to the larger binding energy. We know that the
$\epsilon$-phase of Ga\textsubscript{2}O\textsubscript{3} has a relatively high
static dielectric constant.\textsuperscript{27)} There is no theoretical
report suggesting the large effective mass of the electron and hole. In
general, the layer-type crystalline structures tend to have large
effective masses, which is not the case here. The spin-orbit splitting
in the valence band is frequently present in many semiconductors with
long search history. Our experimental results do not find themselves the
splitting-related distinct structures. Theoretical studies on the
electronic structures have not discussed the splitting in this phase.
Even worse, an enlarged view of the band diagram near the valence band
is not available. Therefore, theoretical works cannot offer a basis for
this splitting effect to analyze our experimental data. Nevertheless, a
weak inflection point observed near 5.5 eV in the imaginary dielectric
function (\(\epsilon_{2}\)) might be reminiscent of this splitting. The
difference from the fundamental bandgap energy amounts to 0.5 or 0.6 eV,
consistent with those found in many semiconductors.\textsuperscript{39)}
We do not, however, consider this splitting here for the abovementioned
reasons. We evaluate \emph{E}\textsubscript{1},
\emph{A}\textsubscript{1}, and \(\Gamma\)\textsubscript{1} to be 6.74
eV, 0.36, and 110, respectively. The \emph{E}\textsubscript{1}
transition parametrization is complicated because this is outside of the
measured energy range. The related parameters are mutually
interdependent.

\begin{figure}
    \caption{Individual contribution to the imaginary part of the
dielectric function of the two energy gaps for
Ga\textsubscript{2}O\textsubscript{3}. They are obtained from Eqs. (1)
and (2) for the \emph{M}\textsubscript{0} contribution, and from Eq. (3)
for the \emph{E}\textsubscript{1}-gap contribution.}
\includegraphics[width=4.99724in,height=3.3315in]{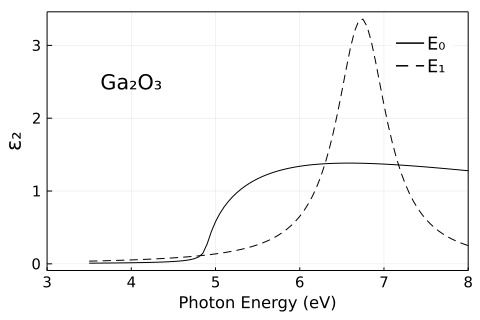}
\end{figure}

\begin{figure}
    \caption{Individual contribution to the imaginary part of the
dielectric function of the two energy gaps for
Ga\textsubscript{2}O\textsubscript{3}.}
\includegraphics[width=4.99724in,height=3.3315in]{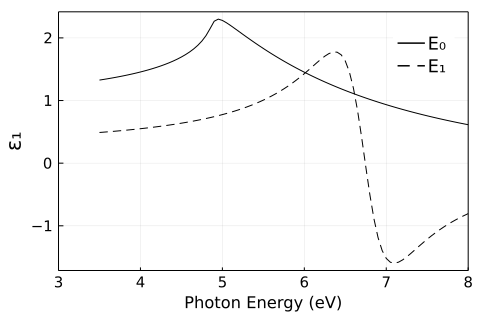}
\end{figure}

\begin{figure}
    \caption{Temperature dependences of direct bandgap
\emph{E}\textsubscript{0} observed in
Ga\textsubscript{2}O\textsubscript{3} (circles). A curve (solid)
corresponds to a numerical fit to the experimental data field using Eq.
(4) with phonon dispersion-related parameters.}
\includegraphics[width=4.49881in,height=5.0126in]{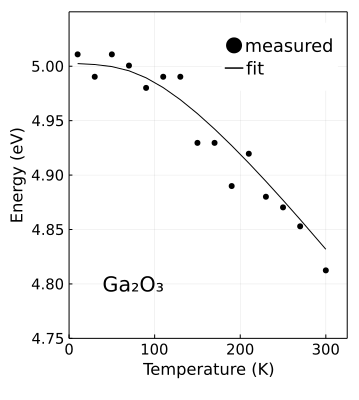}
\end{figure}

\hypertarget{fitting-with-puxe4sslers-model}{%
\subsection{3.2 Fitting with P\"assler's
model}\label{fitting-with-puxe4sslers-model}}

Closed circles in Fig. 4 show the temperature dependence of the bandgap
energy (TDBGE) in Ga\textsubscript{2}O\textsubscript{3}. One confirms
that the bandgap energy (\(E_{0}\)) is a monotonically decreasing
function of temperature (\emph{T}).\textsuperscript{28)} Because the
repulsion between electrons decides the bandgap energy, thermal
expansion of the crystal volume will result in the bandgap shrinkage. As
earlier mentioned, the interface-strain-effect-induced energetic shift
is negligible in our case. The external application of stress gives rise
to a change, a typical magnitude of which is amounted much smaller than
the temperature-induced energetic shift observed here. The interface
strain is several orders of magnitude smaller than the typical strain
values of these external stress experiments.

Usually, one analyzes the TDBGE using Varshni's
model.\textsuperscript{41)} This model assumes an unreasonably large
phonon width. Therefore, the difficulty is often associated with the
interpretation of the fitting parameters. Instead of that model, we use
P\"assler's model to fit the experimental data\textsuperscript{29),30)}.
Here the corresponding equation as a function of temperature can be
expressed:\textsuperscript{31)}

\begin{equation}
E_{0}\left( T \right) = E_{0}\left( 0 \right) - \alpha \Theta \frac{\left( 1 - 3\Delta^{2} \right)}{\exp \left( \frac{\Theta}{T} \right) - 1} + 
\frac{3\Delta^{2}}{2}\left( \sqrt[6]{1 + \frac{\pi^{2}}{3\left( 1 + \Delta^{2} \right)}\left( \frac{2T}{\Theta} \right)^{2} 
+ \frac{3\Delta^{2} - 1\ }{4}\left( \frac{2T}{\Theta} \right)^{3} + \frac{8}{3}\left( \frac{2T}{\Theta} \right)^{4} + \left( \frac{2T}{\Theta} \right)^{6}} - 1 \right)
\end{equation}

We here explain the meaning of the related symbols: $\alpha$ is the
high-temperature limit of a slope, $\Delta$ is a phonon-width-related
parameter, $\Theta$ an effective phonon temperature, and \(E_{0}\) a
fundamental gap's energy, respectively. As later discussed in detail, we
fixed $\Delta$ (= 0.35) and $\Theta$ (= 343 K) parameters as decided from PDOS (phonon
density-of-states) spectrum. The determined values for the
abovementioned parameters in Ga\textsubscript{2}O\textsubscript{3} are
as follows: $\alpha=1.0 \times 10^{-3}$ eV/K,
\(E_{0}\left( 0 \right) = 5.00\)~eV, respectively. A solid trace in Fig.
4 corresponds to the fit for the case that the width-related parameter
($\Delta$) is equal to 0.35. The calculated result agreed reasonably with the
experimental monotonical behavior. According to
P\"assler\textsuperscript{29),30)}, a critical temperature range certainly
exists where the curvature in the temperature-dependent energy gap curve
sensitively depends on the materials' properties. One can calculate this
temperature range by using the effective phonon temperature and
dimensionless phonon dispersion coefficient. The definitions of these
two quantities can find themselves later in detail. The critical
temperatures range from \emph{T}\(\)50 K to \emph{T}\(\)160~K for this
polymorph. Our measurement indeed covered this temperature range,
enabling us to evaluate the materials' parameters. Change of the
polymorph gives rise to the changes in this temperature range and the
curvature in the temperature-dependent energy gap curve. Such a
polymorph-dependent comparative study may be intriguing, which is albeit
beyond the scope of this work.

We describe how to determine the effective phonon temperature ($\Theta$) from
the PDOS lineshape. We assume that this temperature ($\Theta$) is equal to an
averaged phonon temperature ($\Theta$\textsubscript{p}). That is:
\(\Theta_{P} = \left( \Theta_{U} + \Theta_{L} \right)/2\). Here,
\(\Theta_{L}\) (\(\Theta_{U}\)) stands for a temperature corresponding
to energetically-lower (-upper) acoustic (optical) phonon's peak
frequency in the theoretically-calculated $\epsilon$-phase
PDOS.\textsuperscript{42)} For the conversion from the peak frequency to
the temperature, we adopted Boltzmann's constant
(\emph{k}\textsubscript{B}). The TDBGE analysis uses the phonon
temperature ($\Theta$\textsubscript{p}) because we know empirically that the
effective phonon temperature ($\Theta$) is almost equal to the effective phonon
temperature ($\Theta$\textsubscript{p}).

Next, let us define a dimensionless dispersion coefficient related to
the averaged variance in the following:

\begin{equation}
\Delta_{P} = \frac{\sqrt{V_{P} - \mu_{P}^{2}}}{\mu_{P}}
\end{equation}

Here, \emph{V}\textsubscript{p} stands for the averaged variance between
those of the acoustic and the optical phonons. $\mu$\textsubscript{p} means
averaged phonon frequency ($\mu$\textsubscript{p} =
$\Theta$\textsubscript{p} /\emph{k}\textsubscript{B}). We evaluate the
coefficient \(\Delta_{P}\) of Ga\textsubscript{2}O\textsubscript{3} to
be \emph{ca.} 0.50. An empirical rule established from a database of a
variety of materials tells us that the width-related parameter ($\Delta$) tends
to be approximately 70\% of \(\Delta_{P}\).\textsuperscript{43)} We used
these values for the above analysis, as shown with the solid line in
Fig. 4.

We comment about the parameters used in the bandgap energy fit {[}Eq.
(4){]} to the experimental data. Because the visual inspection of the
experimental data enables us to determine \(E_{0}\left( 0 \right)\) we
can concentrate on evaluating the slope parameter ($\alpha$) with the fitting.
We could quickly determine an initial guess of parameterization for the
slope ($\alpha$) because this may correspond to the energy difference between
\emph{T} = 10 K and room temperature. After the regression analysis, the
deviation from the initial guess is in a reasonable range. Reducing the
number of fitting parameters is an advantage. Therefore, the
significance of the phonon-property-related discussion is the reduction
of the ambiguity in the analysis.\textsuperscript{31)}

\hypertarget{discussions}{%
\subsection{3.3 Discussions}\label{discussions}}

In the remainder, we compare the dimensionless dispersion coefficient
and Debye's temperature with various
semiconductors\textsuperscript{43),44)}. First, let us define the
following quantity:
\(r_{LU} = \frac{\Theta_{L}}{\Theta_{U}}\). We call it a
relative phonon temperature. Discussion in terms of this quantity gives
rise to better insight. The relative phonon temperature
(\(r_{LU}\)) of Ga\textsubscript{2}O\textsubscript{3} is
0.46. This value is similar to those in SiC or AlN widegap
semiconductors. Open circles and a closed circle in Fig. 5 show the
magnitude of dimensionless dispersion coefficients
(\(\Delta\)\textsubscript{P}) as a function of the relative phonon
temperature (\(r_{LU}\)). We calculated the lower and
upper zone boundaries for \(\Delta\)\textsubscript{P} according to Ref.
44. Solid and dashed traces, respectively, represent these. We used the
following equation to draw these functions:

\begin{equation}
\frac{1 - r_{LU}}{1 + r_{LU}} < \Delta_{P} < \frac{\sqrt{\left( 1 - r_{LU} \right)^{2} 
+ \frac{2}{3}r_{LU}^{2} + 0.02}}{1 + r_{LU}}
\end{equation}

\begin{figure}
    \caption{Visualization (open squares and a closed circle) of the roughly
monotonic decrease of the magnitudes of dimensionless dispersion
coefficients \(\Delta\)\textsubscript{P}, with increasing the relative
phonon temperature, \(r_{LU}\). Observing that the
partial dispersion coefficients for the lower and upper sections are
limited to the intervals for almost all materials readily establishes
the lower and upper boundaries for this ratio.}
\includegraphics[width=6.09724in,height=4.57205in]{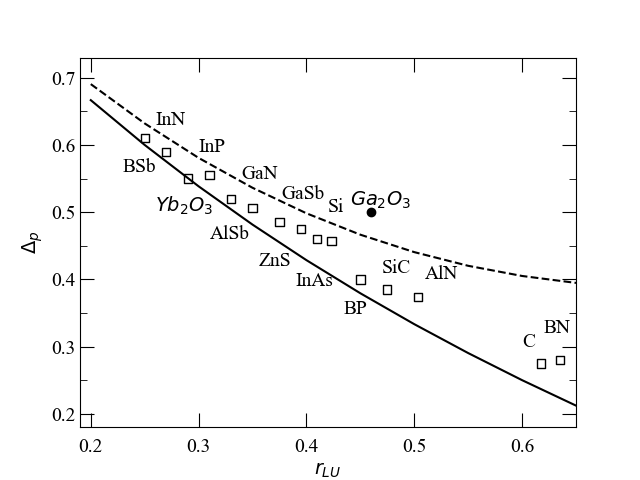}
\end{figure}

\begin{figure}
    \caption{Open squares and a closed circle show the magnitudes of
the ratio \(\frac{_{}\left( \right)}{_{}}\) plotted against the relative
phonon temperature. \(\frac{_{}\left( \right)}{_{}}\)Here,
\(\Theta_{D}\left( \infty \right)\) stands for the limiting Debye
temperatures.}
\includegraphics[width=6.09724in,height=4.57205in]{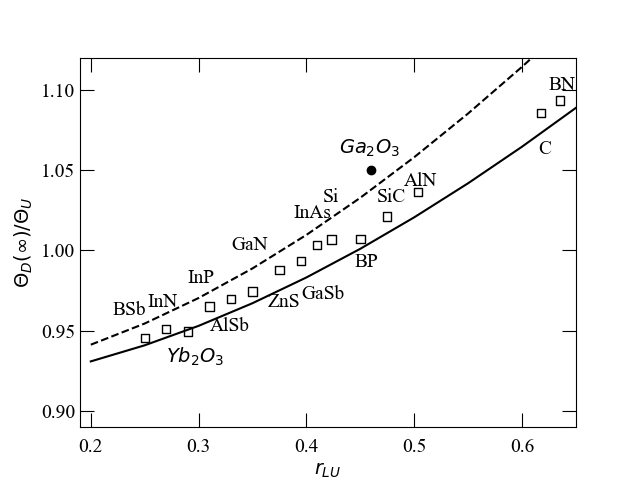}
\end{figure}

The practical limitations of the partial dispersion coefficients of
optical and acoustic phonons determine these boundaries. It is a
well-known fact that many semiconductor materials' phonons exist within
the zones (limits) mentioned above\textsuperscript{44),45)}. As shown in
Fig. 5, we observe that a datum of Ga\textsubscript{2}O\textsubscript{3}
(emphasized using a closed circle) slightly deviates from the upper
boundary. This deviation may be related to the high static dielectric
constant of
$\epsilon$-Ga\textsubscript{2}O\textsubscript{3}\textsuperscript{27)}. Further
studies are necessary to clarify this issue.

Figure 6 shows the relative phonon temperature
(\(r_{LU}\)) dependence of a normalized Debye
temperature \(\Theta_{D}\left( \infty \right)/\Theta_{U}\) (open circles
and a closed circle). Here, \(\Theta_{D}\left( \infty \right)\) stands
for a limiting Debye temperature. The definition of the limiting Debye
temperature is as follows:

\begin{equation}
\Theta_{D}\left( \infty \right) = k_{B}^{- 1}\sqrt{\frac{5}{3}V_{P}}.
\end{equation}

We calculated the normalized Debye temperature
(\(\Theta_{D}\left( \infty \right)/\Theta_{U}\)) for
\(\epsilon\)-Ga\textsubscript{2}O\textsubscript{3} as 1.05. As has been
already explained, the practical limits of the phonon dispersion decide
the upper and lower regions. The resulting equations are as follows:

\begin{equation}
\sqrt{\frac{5}{6}\left( 1 + r_{LU}^{2} \right)} < \frac{\Theta_{D}\left( \infty \right)}{\Theta_{U}} < \sqrt{\frac{5}{6}\left( 1.01 + \frac{4}{3}r_{LU}^{2} \right)}
\end{equation}

\hypertarget{conclusion}{%
\section{4. Conclusion}\label{conclusion}}

We assessed the temperature dependence of the optical properties for
$\epsilon$-Ga\textsubscript{2}O\textsubscript{3} thin films. We have been able to
extract several important material parameters. We interpreted the TDBGE
along a model based on the electron-phonon interaction. We compared the
phonon-related parameters of Ga\textsubscript{2}O\textsubscript{3,} such
as the dimensionless dispersion coefficient of phonons ($\Delta$) and the
normalized Debye temperature
(\(\Theta_{D}\left( \infty \right)/\Theta_{U}\)) with those of many
semiconductors.

\hypertarget{acknowledgments}{%
\section{Acknowledgments}\label{acknowledgments}}

This work was in part conducted in a beamline of the UVSOR facility,
Institute for Molecular Science, Okazaki, Japan. We also acknowledge the
partial financial support of this work from JSPS KAKENHI Grant Number
19K05303. In addition, we acknowledge technical assistance from T. Asai.
T. Nishiwaki, and T. Takeuchi.

\hypertarget{references}{%
\section{References}\label{references}}

1) M. Higashiwaki, H. Murakami, Y. Kumagai and A. Kuramata, Jpn. J.
Appl. Phys. \textbf{55}, 1202A1 (2016).

2) Y.-W. Huan, S.-M. Sun, C.-J. Gu, W.-J. Liu, S.-J. Ding, H.-Y. Yu,
C.-T. Xia and D. W. Zhang, Nanoscale Res. Lett. \textbf{13}, 246 (2018).

3) S. J. Pearton, J. Yang, P. H. Cary, F. Ren, J. Kim, M. J. Tadjer and
M. A. Mastro, Appl. Phys. Rev. \textbf{5}, 011301 (2018).

4) Z. Galazka, Semicond. Sci. \& Technol. \textbf{33}, 113001 (2018).

5) E. G. Villora, K. Shimamura, Y. Yoshikawa, K. Aoki and N. Ichinose,
J. Cryst. Growth \textbf{270}, 420 (2004).

6) H. Aida, K. Nishiguchi, H. Takeda, N. Aota, K. Sunakawa and Y.
Yaguchi, Jpn. J. Appl. Phys. \textbf{47}, 8506 (2008).

7) J. /AAhman, G. Svensson and J. Albertsson, Acta Crystallog. C
\textbf{52}, 1336 (1996).

8) S. Geller, J. Chem. Phys. \textbf{33}, 676 (1960).

9) D. Shinohara and S. Fujita, Jpn. J. Appl. Phys. \textbf{47}, 7311
(2008).

10) K. Kaneko, I. Kakeya, S. Komori and S. Fujita, J. Appl. Phys.
\textbf{113}, 233901 (2013).

11) K. Kaneko, Y. Ito, T. Uchida and S. Fujita, Appl. Phys. Express
\textbf{8}, 095503 (2015).

12) S. Fujita, M. Oda, K. Kaneko and T. Hitora, Jpn. J. Appl. Phys.
\textbf{55}, 1202A3 (2016).

13) R. Roy, V. G. Hill and E. F. Osborn, J. Am. Chem. Soc. \textbf{74},
719 (1952).

14) H. Y. Playford, A. C. Hannon, E. R. Barney and R. I. Walton, Chem.
European J. \textbf{19}, 2803 (2013).

15) Y. Oshima, E. G. V\'illora, Y. Matsushita, S. Yamamoto and K.
Shimamura, J. Appl. Phys. \textbf{118}, 085301 (2015).

16) X. Xia, Y. Chen, Q. Feng, H. Liang, P. Tao, M. Xu and G. Du, Appl.
Phys. Lett. \textbf{108}, 202103 (2016).

17) F. Mezzadri, G. Calestani, F. Boschi, D. Delmonte, M. Bosi and R.
Fornari, Inorganic Chem. \textbf{55}, 12079 (2016).

18) I. Cora, F. Mezzadri, F. Boschi, M. Bosi, M. Caplovicova, G.
Calestani, I. Dodony, B. Pecz and R. Fornari, CrystEngComm \textbf{19},
1509 (2017).

19) Y. Zhuo, Z. Chen, W. Tu, X. Ma, Y. Pei and G. Wang, Appl. Surf. Sci.
\textbf{420}, 802 (2017).

20) D. Tahara, H. Nishinaka, S. Morimoto and M. Yoshimoto, Jpn. J. Appl.
Phys. \textbf{56}, 078004 (2017).

21) H. Sun, K.-H. Li, C. G. T. Castanedo, S. Okur, G. S. Tompa, T.
Salagaj, S. Lopatin, A. Genovese and X. Li, Cryst. Growth \& Design
\textbf{18}, 2370 (2018).

22) D. Tahara, H. Nishinaka, S. Morimoto and M. Yoshimoto, Appl. Phys.
Lett. \textbf{112}, 152102 (2018).

23) H. Nishinaka, N. Miyauchi, D. Tahara, S. Morimoto and M. Yoshimoto,
CrystEngComm \textbf{20}, 1882 (2018).

24) V. Gottschalch, S. Merker, S. Blaurock, M. Kneiss, U. Teschner, M.
Grundmann and H. Krautscheid, J. Cryst. Growth \textbf{510}, 76 (2019).

25) M. Kneiss, A. Hassa, D. Splith, C. Sturm, H. von Wenckstern, T.
Schultz, N. Koch, M. Lorenz and M. Grundmann, APL Materials \textbf{7},
022516 (2019).

26) Y. Xu, J.-H. Park, Z. Yao, C. Wolverton, M. Razeghi, J. Wu and V. P.
Dravid, ACS Appl. Mater. \& Interf. \textbf{11}, 5536 (2019).

27) S. Yusa, D. Oka and T. Fukumura, CrystEngComm \textbf{22}, 381
(2020).

28) H. Pettersson, R. P\"assler, F. Blaschta and H. G. Grimmeiss, J.
Appl. Phys. \textbf{80}, 5312 (1996).

29) R. P\"assler, J. Appl. Phys. \textbf{83}, 3356 (1998).

30) R. P\"assler, Phys. Rev. B \textbf{66}, 085201 (2002).

31) R. P\"assler, Phys. Status Solidi (b) \textbf{236}, 710 (2003).

32) D. Oka, S. Yusa, K. Kimura, A. K. R. Ang, N. Happo, K. Hayashi and
T. Fukumura, Jpn. J. Appl. Phys. \textbf{59}, 010601 (2019).

33) R. E. Denton, R. D. Campbell and S. G. Tomlin, J. Phys. D
\textbf{5}, 852 (1972).

34) J. Furthm\"uller and F. Bechstedt, Phys. Rev. B \textbf{93}, 115204
(2016).

35) J. Kim, D. Tahara, Y. Miura and B. G. Kim, Appl. Phys. Express
\textbf{11}, 061101 (2018).

36) S. Adachi, Phys. Rev. B \textbf{35}, 7454 (1987).

37) S. Adachi, J. Appl. Phys. \textbf{66}, 813 (1989).

38) H. He, R. Orlando, M. A. Blanco, R. Pandey, E. Amzallag, I. Baraille
and M. R\'erat, Phys. Rev. B \textbf{74}, 195123 (2006).

39) S. Adachi and T. Taguchi, Phys. Rev. B \textbf{43}, 9569 (1991).

40) A. R. Go\~ni, K. Syassen, Y. Zhang, K. Ploog, A. Cantarero and A.
Cros, Phys. Rev. B \textbf{45}, 6809 (1992).

41) Y. P. Varshni, Physica (Utrecht) \textbf{34}, 149 (1967).

42) S. Yoshioka, H. Hayashi, A. Kuwabara, F. Oba, K. Matsunaga and I.
Tanaka, J. Phys.: Cond. Matter \textbf{19}, 346211 (2007).

43) R. P\"assler, Phys. Status Solidi (b) \textbf{243}, 2719 (2006).

44) R. P\"assler, J. Appl. Phys. \textbf{101}, 093513 (2007).

45) T. Makino, T. Asai, T. Takeuchi, K. Kaminaga, D. Oka and T.
Fukumura, Jpn. J. Appl. Phys. \textbf{59}, SCCB13 (2020).
\end{document}